\documentclass[aps,pre,superscriptaddress,twocolumn,floatfix]{revtex4-1}
\usepackage{graphicx,amssymb,amsmath,amsfonts,bm,bbold}
\DeclareMathOperator{\sech}{sech}

\begin{document}

\title{Single-qubit remote manipulation by magnetic solitons}

\author{Alessandro Cuccoli}
\affiliation{Dipartimento di Fisica e Astronomia, Universit\`a di Firenze,\\
	via G.~Sansone 1, I-50019 Sesto Fiorentino (FI), Italy}
\affiliation{Istituto Nazionale di Fisica Nucleare, Sezione di Firenze,\\
	via G.~Sansone 1, I-50019 Sesto Fiorentino (FI), Italy}

\author{Davide Nuzzi}
\affiliation{Dipartimento di Fisica e Astronomia, Universit\`a di Firenze,\\
	via G.~Sansone 1, I-50019 Sesto Fiorentino (FI), Italy}
\affiliation{Istituto Nazionale di Fisica Nucleare, Sezione di Firenze,\\
	via G.~Sansone 1, I-50019 Sesto Fiorentino (FI), Italy}

\author{Ruggero Vaia}
\affiliation{Istituto dei Sistemi Complessi, Consiglio Nazionale delle Ricerche,
	via Madonna del Piano 10, I-50019 Sesto Fiorentino, Italy}
\affiliation{Istituto Nazionale di Fisica Nucleare, Sezione di Firenze,
	I-50019 Sesto Fiorentino, Italy}

\author{Paola Verrucchi}
\affiliation{Istituto dei Sistemi Complessi, Consiglio Nazionale delle Ricerche,
	via Madonna del Piano 10, I-50019 Sesto Fiorentino, Italy}
\affiliation{Dipartimento di Fisica e Astronomia, Universit\`a di Firenze,\\
	via G.~Sansone 1, I-50019 Sesto Fiorentino (FI), Italy}
\affiliation{Istituto Nazionale di Fisica Nucleare, Sezione di Firenze,\\
	via G.~Sansone 1, I-50019 Sesto Fiorentino (FI), Italy}

\begin{abstract}
Magnetic solitons can constitute a means for manipulating qubits from a distance. This would overcome the necessity of directly applying selective magnetic fields, which is unfeasible in the case of a matrix of qubits embedded in a solid-state quantum device. If the latter contained one-dimensional Heisenberg spin chains coupled to each qubit, one can originate a soliton in a selected chain by applying a time-dependent field at one end of it, far from the qubits. The generation of realistic solitons has been simulated. When a suitable soliton passes by, the coupled qubit undergoes nontrivial operations, even in the presence of moderate thermal noise.
\end{abstract}

\maketitle

\section{Introduction}
\label{s.intro}

The quest for the most suitable physical architecture for the realization of quantum computing machines attracted an increasing attention in the last two decades. Among the basic requirements that any qubit implementation must fulfill there is the possibility to let the qubit interact with the external world in order to initialize, manipulate and eventually read its quantum state; at the same time, during the actual quantum calculation the quantum register has to be protected from external actions, while the proper, coherent dynamics of the qubit(s) takes place undisturbed\cite{LossD-V98}. The initialization step implies the ability to individually address and manipulate each single qubit of the quantum register, hopefully being able to bring the qubit in the desired, final quantum state by the most direct way and in the shortest interval of time.
When qubits are represented by spin-$\frac12$ localized objects~\cite{cite1,PlaTDLMZJDM2013}
(atoms, molecules, quantum dots,...) their control~\cite{EngelKLM2004} is usually assumed to be obtained by applying suitable sequences of external magnetic fields. Indeed, any unitary action on a single qubit can be represented in terms of a Zeeman interaction lasting for a precise time interval. If the magnetic qubits are embedded in crystalline matrices, or deposited on layered substrates, one has at least to face the following how-tos: {\it{i}}) address the specific qubit upon which the field should act, {\it{ii}}) keep the Zeeman interaction on just for the required interval of time, {\it{iii}}) avoid the apparatus generating the field, and the field itself, to cause disturbing effects on the surrounding qubits. The first two points could be implemented by means of magnetic tips on top of the register so as to act on each qubit; however, such an approach conflicts with point {\it{iii}}), denying the possibility to increase the density of qubits per unit area using currently available technological capabilities. A different scheme is that where some kind of "wires", one for each qubit, depart from a control apparatus and convey to the qubits a suitable signal to accomplish  the assigned task: a practical realization of such scheme can be found, e.g., in nanofabricated microwave transmission lines recently used to manipulate electron and nuclear spin of phosphorus donors in silicon~\cite{PlaTDLMZJDM2013},
and previuosly to control superconducting qubits\cite{Vion02_BlaisWallraff04_Mallet09}.
In this paper we will focus upon the latter scheme, our aim being that of putting forward a proposal for the realization of a magnetic functional wire and an efficient signal. The wire will be modeled as a one-dimensional magnetic array, i.e., a spin-chain; a control apparatus is placed at one end of the chain far enough from the quantum register to avoid any disturbance, bar that mediated by the wire. The wire needs not being quantum, as it only has to carry classical signals acting on the associated qubit as a highly localized magnetic field. The signals could in principle be any arbitrary pulses generated by the control apparatus, but this choice would lead to face the problem of distortion and attenuation, being spin-chains dispersive media, additionally affected by possible noise or imperfections. It is more promising to choose those dynamical magnetic excitations known to be localized in space and time and robust against perturbations, namely solitons.

\section{Solitons in the Heisenberg chain}
\label{s.Soliton}

`Solitons' are nonlinear excitations of a field defined on a one-dimensional space. They travel at constant velocity with a localized profile and are exceptionally stable, as they arise as analytical solutions of the nonlinear equations of motion, when looking for a constant-shape field of the form $\psi(x{-}vt)$. Soliton-like excitations also characterize several spin-chain models~\cite{LongB1979,Fogedby1980,ElstnerM1989,KosevichIK1990,SchmidtSL2011} that support genuine solitons in their continuum approximation. Experiments on real compounds, known to be properly described as spin-$S$ chains, proved that solitons do indeed characterize the energy spectrum, and hence the dynamics and the thermodynamics, of one-dimensional magnetic systems~\cite{KjemsS1978,BoucherRRRBS1980,RamirezW1982,MikeskaS1991}. Therefore it is reasonable to assume that real spin chains, in spite of their being discrete and quantum (the classical limit occurs for $S\,{\to}\,\infty$), still support solitons, as confirmed also by numerical work~\cite{WoellertH2012}. As for quanticity, it has been shown that a semiclassical approach~\cite{CTVV,CGTVV1995} consisting in studying a classical model with $S$-dependent renormalized parameters is appropriate for describing the behavior of quantum spin models with $S\,{>}\,1/2$. Hence, the analysis of soliton generation and propagation, i.e., the part of our scheme that concerns the signal's conveyance to the qubit, is here treated classically.

We consider a classical spin-$S$ Heisenberg chain, i.e., a one-dimensional array of (spin-)vectors $\bm{S}_n\equiv{S}\,\bm{s}_n$, whose magnitude $S$ has the dimension of an action. The unit vectors $\bm{s}_n$ are naturally parameterized by polar coordinates, $\bm{s}_n\equiv(\sin\theta_n\cos\varphi_n,\sin\theta_n\sin\varphi_n, \cos\theta_n)$, with $\varphi_n$ and $\cos\theta_n$ canonically conjugated variables, $\{\varphi_n,\cos\theta_m\}=S^{-1}\,\delta_{nm}$. The Hamiltonian is
\begin{equation}
{\cal H}_{\rm ch}= -JS^2 {\sum}_n \bm{s}_n{\cdot}\bm{s}_{n+1}
  -\gamma S\bm{H}{\cdot}{\sum}_n \bm{s}_n~,
\label{e.Hdiscr}
\end{equation}
where $J\,{>}\,0$ is the ferromagnetic nearest-neighbor exchange, $\bm{H}=(0,0,H)$ is an external magnetic field, $\gamma$ is the gyromagnetic ratio. In the equations of motion, $\partial_t\bm{s}_n=JS\,\bm{s}_n\times(\bm{s}_{n+1}{+}\bm{s}_{n-1}+\bm{h})$,
the frequency scale is set by $JS$, while the dimensionless Zeeman field $\bm{h}\equiv\gamma\bm{H}/(JS)$ can be considered a `tunable' parameter.

\begin{figure*}
	\includegraphics[width=0.5\textwidth]{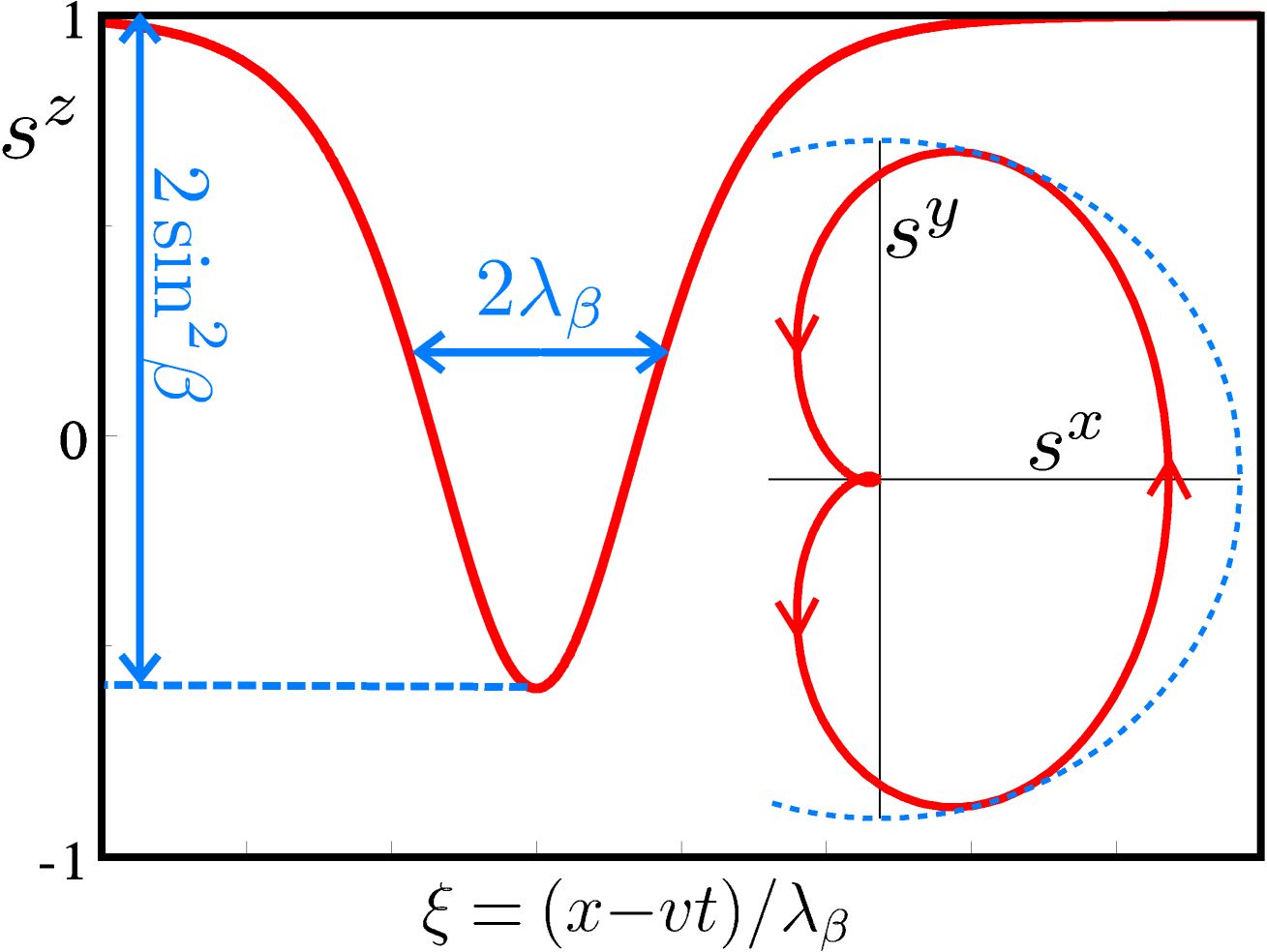}
	~~~~\includegraphics[width=0.4\textwidth]{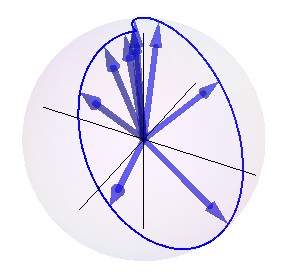}
	\caption{TW soliton for $\tan\!\beta\,{=}\,2$: left panel, $z$-component of the spins, $s^z=\cos\theta_\beta(\xi)$ and (inset) the corresponding trajectory of the in-plane components $s^x$ and $s^y$; right panel, evolution of the spin vector $\bm{s}_\beta(\xi)$, reported at constant intervals $\Delta\xi\,{=}\,0.8$.}
	\label{f.soliton}
\end{figure*}
The low-energy excitations of the Heisenberg chain~\eqref{e.Hdiscr} are spin waves (or `magnons'), but in the continuum approximation (lattice spacing $d\,{\to}\,0$) Tjon and Wright~\cite{TjonW1977} (TW) found an analytical `one-soliton' solution and numerically studied multiple-soliton excitations, showing that solitons are stable under collisions, a property confirmed by the inverse-scattering method~\cite{Takhtajan1977}. The general form of the TW soliton is
\begin{equation}
\left\{
\begin{aligned}
 \theta_{\beta} &= 2\sin^{-1}(\sin\!\beta\,\sech\xi) ~,
\\
\varphi_{\beta} &= \varphi_0 + \cot\!\beta\,\xi
+\tan^{-1}(\tan\!\beta\,\tanh\xi) ~,
\end{aligned}
\right.
\label{e.twsol}
\end{equation}
where
$\xi\equiv\,(x{-}vt)/\lambda_{\beta}$ and $x\,{=}\,nd$ is the `continuum' coordinate. The soliton \emph{amplitude}, characterized by the dimensionless parameter $\beta\in(0,\pi/2)$ ($\theta\,{\le}\,2\beta$) is related with the velocity,
$\cos\!\beta=v/({2dJS\sqrt{h}})$~, 
and determines the soliton \emph{length}~
$\lambda_{\beta}=d/({\sqrt{h}\sin\!\beta})$~, 
\emph{time scale}~
$\tau_{\beta}=(JS\,h\,\sin\!2\beta)^{-1}$~, 
and {\em energy}~
$\varepsilon_{\beta}=8JS^2\sqrt{h}\,\sin\!\beta$~. 
Fig.~\ref{f.soliton} reports a typical TW soliton; note that solitons with larger amplitude $\beta$ have larger energy ($\sim\sin\!\beta$), are narrower $(\sim1/\sin\!\beta)$ and slower ($\sim\cos\!\beta$). 
Although there are no known analytic soliton solutions of the discrete model, the continuum approximation holds for configurations that vary slowly on the scale of the lattice spacing $d$, so that the solution~\eqref{e.twsol} approximately applies also to the chain model~\eqref{e.Hdiscr} provided that $\lambda_{\beta}\gg{d}$, i.e., $\sqrt{h}~\sin\!\beta\,{\ll}\,1$. This is generally true in real systems, whose typical exchange energies are of the order of tenths-hundreds of Kelvin degrees: as $\mu_{\rm{B}}=0.67$~K/Tesla, only very large fields could break the inequality.

\section{How to inject a soliton in the chain?}
\label{s.inject}

A soliton can be created by applying a time-dependent magnetic field at one end of the chain. If the field is localized to an extension smaller than that of the soliton to be generated, the process can be schematized as if a suitable field $\bm{b}(t)$ be applied just on the first spin, say $\bm{S}_{1}$. This is equivalent to solve the equations of motion for a chain with an extra spin $\bm{S}_{0}(t)=\gamma\bm{b}(t)/J$ whose time behavior is enforced. The time dependence of $\bm{S}_{0}(t)$ must be the same of that induced by a soliton coming from the region $n<0$ if the chain were (infinitely) extended: indeed, that soliton would continue to travel forward and would be successfully injected into the chain, at least in the continuum limit. Hence the field acting on the chain end is taken as $\bm{b}(t)=\gamma^{-1}JS\,\bm{s}_{\beta}(t)$, where $\bm{s}_{\beta}(t)$ is the TW soliton~\eqref{e.twsol} and $\beta$ can be chosen.

With the above time-dependent constraint we have numerically~\cite{KrechBL1998,Yoshida1990,ForestR1990,Suzuki1992,TsaiLL2005} solved the chain dynamics: it turns out that a `soliton' $\bm{s}_n(t)$ is indeed generated in the {\em discrete} chain. The initial state was taken as the minimum-energy one, i.e., at temperature $T\,{=}\,0$, $\{\bm{s}_n\,{=}\,\hat{\bm{z}},~n\,{\ge}\,1\}$. Further runs, aimed at studying the stability against disorder, started from typical finite-$T$ configurations with correlators $\big\langle(s^x_n{-}s^x_{n+1})^2\big\rangle
=\big\langle(s^y_n{-}s^y_{n+1})^2\big\rangle\simeq{\cal{T}}$ and $\big\langle(s^x_n)^2\big\rangle=\big\langle(s^y_n)^2\big\rangle\simeq{\cal{T}}/\sqrt{h}$, where ${\cal{T}}\equiv{k}_{_{\rm{B}}}T/JS^2$.

\begin{figure*}
	\includegraphics[width=0.9\textwidth]{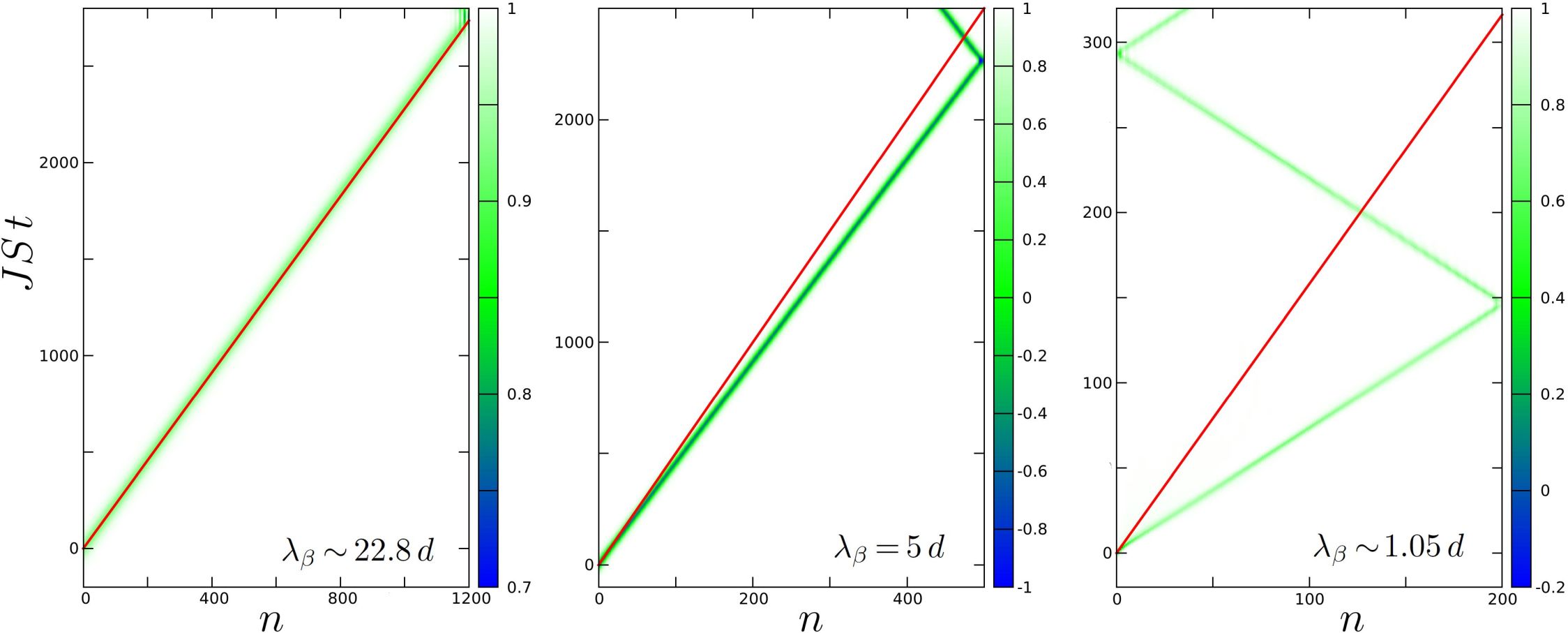}
	\caption{Evolution of $s_n^z(t)$ for solitons generated in discrete chains by injection of TW shapes $\bm{s}_\beta(t)$ with three different lengths:  $\lambda_{\beta}\,{\sim}\,22.8\,d$ ($h\,{=}\,0.05$, $\tan\!\beta\,{=}\,0.2$), $\lambda_{\beta}\,{=}\,5\,d$ ($h\,{=}\,0.05$, $\tan\!\beta\,{=}\,2$), $\lambda_{\beta}\,{\sim}\,1.05\,d$ ($h\,{=}\,1$, $\tan\!\beta\,{=}\,3$). The white-to-green-to-blue shading corresponds to the value of 
	$s^z_n(t)$, visually displaying the progressive deviation of the spin vector from the $z$-direction, and permits to appreciate how the excitation is localized in space and time. The red lines' slope corresponds to the velocity of the injected TW soliton; and reflection occurs at the open chain end.}
	\label{f.gensol}
\end{figure*}

\begin{figure*}
	\includegraphics[width=0.9\textwidth]{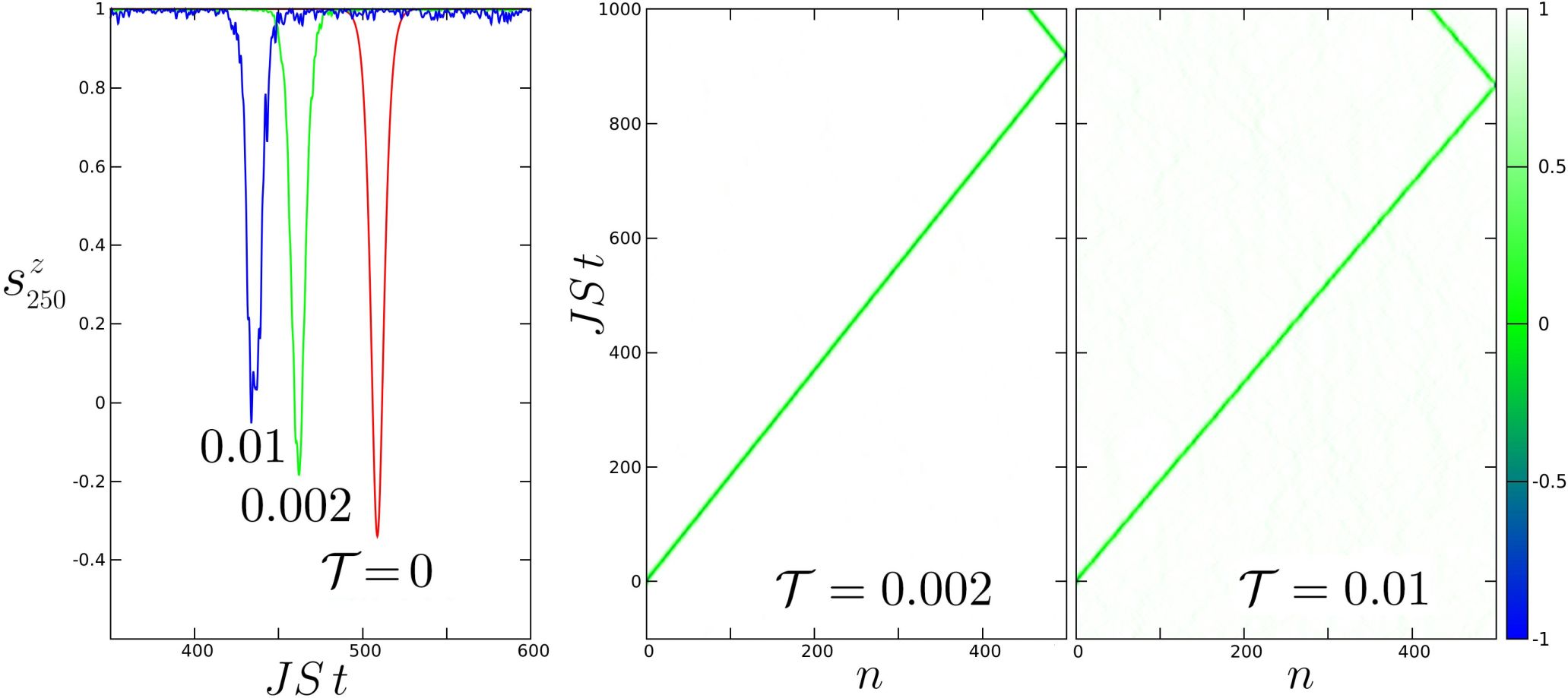}
	\caption{Solitons generated in a discrete chain of $N\,{=}\,500$ spins by injecting a TW shape $\bm{s}_\beta(t)$ with $h\,{=}\,0.2$ and $\tan\!\beta\,{=}\,2$. Left panel: $s_{250}^z(t)$ for ${\cal T}=0$ (red line), ${\cal T}=0.002$ (green line), ${\cal T}=0.01$ (blue line). Density plots: evolution of $s_n^z(t)$ for two different temperatures with shading as in Fig.~\ref{f.gensol}.}
	\label{f.gensol_temp}
\end{figure*}

The generated solitons are localized, stable, and move with constant velocity as displayed in Fig.~\ref{f.gensol}, where their velocity is compared with that of the corresponding injected TW soliton. From the observed velocity it is possible to estimate an effective parameter $\beta'$, which happens to agree with the fit of the shape using Eq.~\eqref{e.twsol}, and the calculated total work made by the forcing term agrees with $\varepsilon_{\beta'}$, the energy of the TW-like soliton. As $\beta'\,{<}\,\beta$ the generated soliton has slightly less energy and is faster than the theoretically injected one: the difference is more evident the narrower the injected soliton, and is due to the discreteness of the spin chain. Also for nonzero temperature solitons are generated: while their features depend on the value of ${\cal T}$, their dynamics is not spoiled by the underlying noise, as shown in Fig.~\ref{f.gensol_temp}. In conclusion, discrete spin chains support propagating solitons similar to the known analytical solutions of the continuum model; these can be generated by applying a suitable time-dependent localized magnetic field at one end of the chain and are stable against thermal noise.

Typical values for the exchange interaction in common quasi one-dimen\-sional magnets are $JS^2\,{\sim}\,1\div10^3$\,K, and the applied field has to be such that $h\,{\ll}\,1$; hence, taking $h\,{=}\,0.05$ and $S=\hbar$ (i.e., a spin-1 chain) one finds a physical estimate of the soliton velocity $v\,{\sim}\,10^{11}\div10^{14}$ lattice spacings per second. For a chain of $10^3$ spins, this amounts to traveling times of the order of $10\div10^{-2}$ ns, which are  
adequate, given the attainable qubit coherence times.

\section{Any single qubit is coupled to its chain}
\label{s.GettingThrough}

\begin{figure*}
	\includegraphics[width=0.9\textwidth]{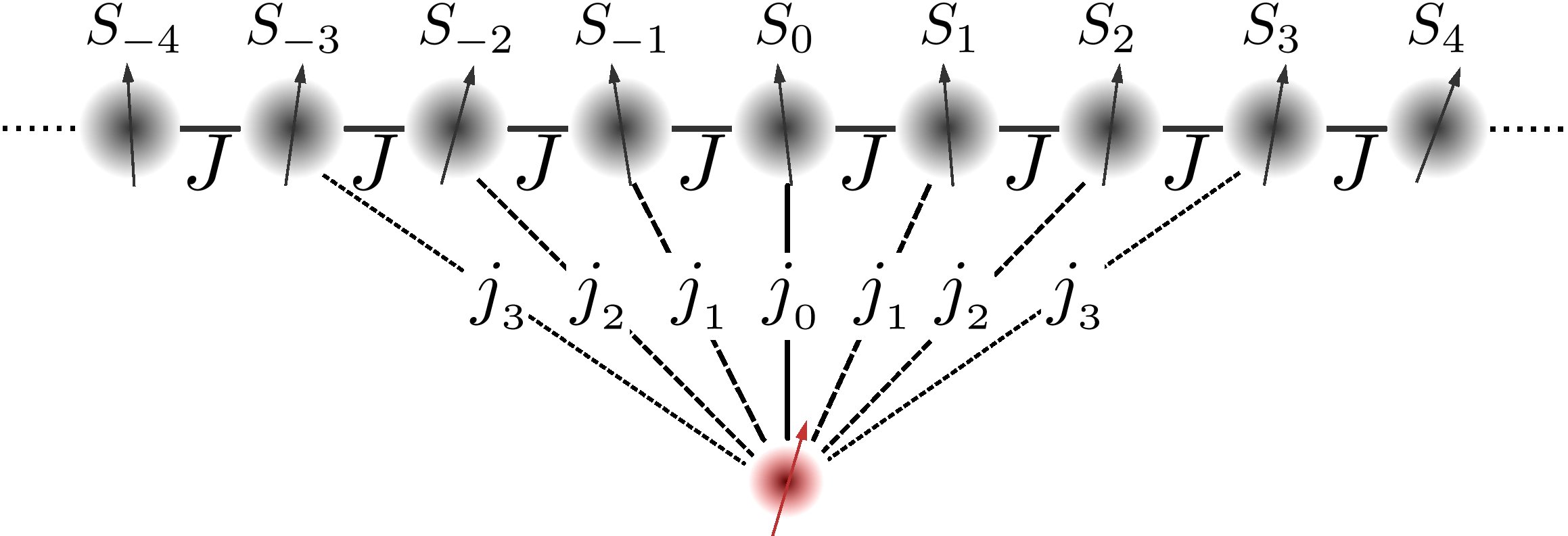}
	\caption{Physical realization of the scheme of Eq.~\eqref{e.Htotrange}: a qubit interacts with a bunch of moments of a classical spin-chain, with couplings $j_n=g\,p_n$.}
	\label{f.qubit-chain}
\end{figure*}

The qubit, here taken as spin-$\frac12$ particle, is close to the site labeled by $n\,{=}\,0$ of a given spin chain as depicted in Fig.~\ref{f.qubit-chain}, so the dynamics of the qubit's spin $\hbar\hat{\bm{\sigma}}/2$ is ruled  by the Hamiltonian
\begin{equation}
\hat{\cal{H}}_{\rm qubit} = -\frac{\hbar\bm{\hat\sigma}}2
 \cdot\Big(\gamma_\sigma\bm{H}+ g S\sum_n p_n\bm{s}_n\Big) ~,
\label{e.Htotrange}
\end{equation}
where $\gamma_\sigma$ is the qubit's gyromagnetic ratio. The isotropic exchange couplings $j_n\,{\equiv}\,g\,p_n$, with $\sum_n{p_n}\,{=}\,1$, such that $g$ represents the overall intensity of the qubit-chain interaction, rapidly decrease with the distance $|n|$. A soliton passing by will affect the state of the qubit~\cite{CNVV2014}, without disturbing its environment, e.g., other qubits that may sit nearby but are not coupled to the same chain.

 \begin{figure*}
 	\includegraphics[width=0.9\textwidth]{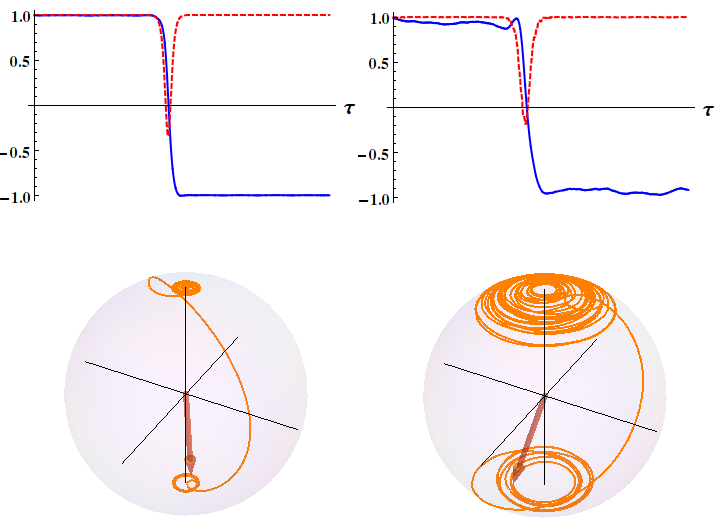}
 	\caption{Top panels: Time evolution of $a^z(\tau)$ (solid blue line) for the qubit interacting with the soliton (dashed red line) generated by injecting TW-shape  ($\tan\!\beta\,{=}\,2$, $h=0.2$) with ${\cal T}=0$ (left panel) and ${\cal T}=0.002$ (right panel); $\mu\,{=}\,1$, $\delta\,{=}\,1$, and $\alpha\,{=}\,0$. Bottom panels: Parametric plots of the qubit's state evolution on the Bloch sphere, under the same conditions of the respective upper panel.}
 	\label{f.tb20_gen}
 \end{figure*}

Note that the chain dynamics is practically insensitive to the qubit, since the maximal qubit's energy gain (occurring for a complete flip), $\delta{E}\,{=}\,\hbar(gS{+}\gamma_\sigma{H})$, is much smaller than the soliton energy $\varepsilon_{\beta}$; indeed,
\begin{equation}
\frac{\delta{E}}{\varepsilon_{\beta}}
=\frac{gS{+}\gamma_\sigma{H}}{8\gamma{H}}
\,\frac{\hbar}{S}\,\frac{\sqrt{h}}{\sin\!\beta} ~\ll~1
\label{e.noback}
\end{equation}
because of the validity of the criterion $\sqrt{h}\,{\ll}\sin\!\beta$ and of the assumption that the chain spins be almost classical, $S\,{>}\,\hbar$.

Let us then study the change of the qubit's state from `before' to `after' a given soliton travels along the chain. The qubit's density matrix $\hat\rho(\tau)$, being $\tau\equiv\gamma{H}\,t=JSh\,t$ the dimensionless time, can be represented by the normalized vector $\bm{a}(\tau)\equiv{\rm{Tr}}\big[\hat\rho(\tau)\,\bm{\hat\sigma}\big]$, whose time evolution obeys the equation of motion
\begin{equation}
 \partial_\tau\bm{a} = 
 \bm{a}\times\big[\delta\,\hat{\bm{z}}+\mu\,\tilde{\bm{s}}(\tau)\big] ~,
 \label{e.dtaua}
\end{equation}
with $\hat{\bm{z}}=(0,0,1)$; the dimensionless parameters
\begin{equation}
\delta=\frac{\gamma_\sigma}{\gamma}
~,~~~~~~
\mu \equiv \frac{gS}{\gamma{H}}
\label{e.deltamu}
\end{equation}
characterize the qubit's interactions with the field and with the chain, while
\begin{equation}
 {\bm{\tilde{s}}}(\tau) = {\sum}_np_n\,{\bm{s}}_n(\tau) 
\label{e.stildedisc}
\end{equation}
can be seen as a `smeared' soliton shape arising from the last term of the Hamiltonian~\eqref{e.Htotrange}. As said above, the injected discrete soliton is close to ideal TW shape~\eqref{e.twsol}, namely ${\bm{s}}_n(\tau)\simeq\bm{s}_\beta\big(n\sqrt{h}\sin\!\beta-\tau\sin\!2\beta\big)$.
The soliton effective field is not normalized, $|\bm{\tilde{s}}(\tau)|<1$, except when $p_n\,{=}\,\delta_{n0}$ and ${\bm{\tilde{s}}}(\tau)=\bm{s}_0(\tau)$, i.e., the qubit just interacts with the single spin $\bm{s}_0$~\cite{CNVV2014}. The kind of coupling between the qubit and the chain spins rules the dependence on $n$ of the $p_n$'s: for instance, an exchange coupling is likely to decay exponentially, while a dipolar or RKKY coupling would give a power-law, $p_n\,{\sim}\,|n|^{-\kappa}$ with $\kappa=3$; as long as the $p_n$ are summable ($\kappa\,{>}\,1$) the discrete convolution~\eqref{e.stildedisc} is well defined and what matters is the 'width'  $\alpha$ of the distribution of the $p_n$. Therefore, for the present purpose, we can choose a simple Gaussian,  $p_n=Ae^{-n^2/2\alpha^2}$, such that $\alpha$ characterizes the interaction range in units of $d$. Equations~\eqref{e.dtaua} with ${\bm{\tilde{s}}}(\tau)$ from Eq.~\eqref{e.stildedisc} have to be integrated numerically. When the soliton is far, one has $\tilde{\bm{s}}(\tau)\,{=}\,\hat{\bm{z}}$, so that well before and after the soliton transit the qubit precesses uniformly around $\hat{\bm{z}}$ with frequency $\omega\,{=}\,\mu{+}\delta$. In order to isolate the qubit evolution exclusively due to the soliton transit, it is convenient to choose a polarized initial (at some $\tau\,{\ll}\,{-1}$) state, $\bm{a}_{\rm{in}}\,{=}\,\hat{\bm{z}}$. The experimental configuration sets the input parameters $\mu$, $\delta$, $\alpha$, and the reduced field $h$, while the soliton-shape parameter $\beta$ can be varied. 

Being the process aimed at `writing' on the input quantum register, by suitably adjusting the parameters $\mu$, $\delta$, $\alpha$, and $\beta$, one has to be able to obtain asymptotic states $\hat\rho_{\rm{out}}\equiv\hat\rho(\tau{\gg}1)$ maximally different from the initial one. The general asymptotic behavior of the final qubit state is
$\bm{a}_{\rm{out}}\,{=}\,\big[a^\perp_{\rm{out}}\,\cos(\omega\tau{+}\phi),
a^\perp_{\rm{out}}\,\sin(\omega\tau{+}\phi),a^z_{\rm{out}}\big]$
with $a^\perp_{\rm{out}}=\sqrt{1{-}(a^z_{\rm{out}})^2}$ and $\phi$ a constant. For $\alpha\,{=}\,0$ and ideal TW solitons it is shown in Refs.~\cite{CNVV2014} that in the ideal case $\delta\,{=}\,\mu\,{=}\,1$ the qubit is fully flipped by the soliton, $a^z_{\rm{out}}=-1$, irrespective of the value of $\beta$; however, in the case of thermal noise and/or injected solitons, the choice of $\beta$ turns out to affect the qubit's response and can be used to optimize it~\cite{CNVV2015}.

In Fig.~\ref{f.tb20_gen} we show the qubit state evolution upon injection of a soliton with $\tan\!\beta\,{=}\,2$, of characteristic length  $\lambda_\beta\,{=}\,5d$, in the case of point-like ($\alpha\,{=}\,0$) interaction with the qubit: in the upper panels one can appreciate how the evolution of the qubit's $z$-component $a^z(\tau)$ follows that of the passing generated soliton, both for zero (left) and finite  (right) temperature; the bottom panels display the overall trajectory of the qubit's magnetization on the Bloch sphere. The qubit's behavior under the action of a generated soliton is similar to that for ideal solitons~\cite{CNVV2015}; in particular, for ${\cal{T}}\,{=}\,0$ an almost complete flip is obtained. Of course, for ${\cal{T}}\,{>}\,0$ the asymptotic value $a^z_{out}$ fluctuates under the thermal fluctuations of the spin chain; nevertheless, the qubit evolution stays \textit{on} the Bloch sphere and there is no decoherence, as the (noisy) effective field acting on the qubit is still classical.

\section{Conclusions}

In conclusion, we have shown that properly acting by an external field at one end of a magnetic chain, solitonic excitations can be successfully injected and propagate without relevant distortion along a magnetic chain. The discreteness effects decrease when the soliton length is larger. The propagation of such solitons is relatively robust against thermal noise up to $T\,{\sim}\,0.01\,JS^2/{k}_{_{\rm{B}}}$. 
This generation has been achieved by the application of a boundary term representing a proper time-dependent magnetic field which only affects the first spin of the chain. In a realistic situation this forcing field would enter the chain with a characteristic penetration lenght $\ell$. For this reason we are currently investigating the chain dynamics in presence of a boundary term with such a spatial decay. Preliminary results show that in such situation solitons are successfully generated even by using magnetic fields whose intensity is reduced of a factor $\sim d/\ell$.

The injected soliton is able to flip a qubit significantly far from the chain edge,  assuring that the qubits possibly embedded in the same solid-state matrix are not directly affected by the external field needed to generate that soliton.
When physical realizations of magnetic chains are considered, magnetic anisotropies are usually observed; it will therefore be important to investigate how they may affect the scheme here proposed. We defer such investigation to future work, but in the meantime it is useful to remind that solitonic excitations are sustained also in anisotropic chains~\cite{LongB1979,KosevichIK1990}, so that our scheme should continue to work also for anisotropic compounds.

\end{document}